\documentclass[11pt]{article}
\usepackage{moriond,epsfig}

\def\be{\begin{equation}}
\def\ee{\end{equation}}
\def\bea{\begin{eqnarray}}
\def\eea{\end{eqnarray}}

\begin{document}
\vspace*{4cm}
\title{MEASUREMENTS OF $|V_{ub}|$ AT BELLE
}

\author{ A. SUGIYAMA \\
( for Belle collaboration )}

\address{Department of Physics, Saga University,\\ 
1 Honjyou-cyou, Saga, 840-8502, Japan}

\maketitle\abstracts{
We have tried to measure $|V_{ub}|$ using  inclusive semileptonic
decay measurements with three different methods.
The full reconstruction tag is the one of the most feasible method
to measure inclusive process. 
%We just demonstrate its ability by showing
%results on $|V_{cb}|$ process.
Semileptonic $B\rightarrow D^{(*)}\ell\nu$ tag is another  method 
to separate signal and tag  $B$ mesons. 
New kinematic variable
is used to enhance double semileptonic decays of both $B$ meson.
We also improve the neutrino reconstruction technique 
solving a combination of  decay products into signal and tag 
$B$ mesons 
using the simulated annealing method.
}

\section{Introduction}

% status of Vub measurements
The precise measurements of $|V_{ub}|$ is one of the  potential
measurements to show the first discrepancy of
KM-schemed  standard model~\cite{KM}.
However current $|V_{ub}|$ measurements are limited by statistics,
model related systematics and theoretical uncertainties.
B factory experiments have an opportunity to overcome statistical
limitation easily and
the latter two uncertainties are main theme of the
analysis.
Exclusive analysis has  introduced inevitable
 theoretical uncertainties on hadronic part
though the analysis is simpler.
Contrary inclusive analysis has
difficulties due to ambiguous kinematics
but has a chance to reduce theoretical uncertainty,
if larger kinematic region could be measured.
Theorists propose formula to extract $|V_{ub}|$ from a partial branching
fraction well defined kinematic region using the Operator Product
Expansion(OPE) based on  Heavy Quark Effective Theory(HQET) or Heavy
Quark Expansion(HQE) method~\cite{OPE}.
Measurements of the shape function,
hadronic mass moments and lepton spectra moments
may help to reduce theoretical uncertainties.

%experiment
From experimental point of view, the difficulties of  inclusive studies
are based on a lack of knowledge related to
(1)whether event is $B\bar B$ or continuum,
(2)which particles are decay daughters of either $B$ meson,
(3)whether the signal $B$ decay into $u\ell\nu$ or $c\ell\nu$
(4)ambiguity of kinematics.
The neutrino reconstruction technique can solve the last issue
with an assumption of all produced particles are observed except 
neutrino in
the final state, though detection efficiency will be sacrificed.
The complete full reconstruction of the opposite $B$
meson in hadronic decay mode
must be the best solution for the issue (1) and (2).
The combination of two techniques is an ideal method to study
$B\rightarrow X_u \ell \nu$ and it is equivalent to the complete
reconstruction of two $B$ meson from $\Upsilon(4S)$.
However the efficiency of
the full reconstruction method
is not so large due to variety of $B$ meson decays in hadronic modes.
Belle has tried to develop  other new methods 
to study inclusive $b\rightarrow u\ell\nu$
decay besides the full-reconstruction method.

We tried to include the semileptonic decay modes
as a $B$ meson tag as they are well established and has larger
branching fractions as a decay mode.
The semileptonic decay mode for  $B$ meson tag has statistical
advantage against the full reconstruction method
but an introduction of  another neutrino is
inevitable in the final state in addition to $\nu$ from the signal
$b\rightarrow u\ell\nu$.
Using all constraints, kinematics can be solved with a small ambiguity
as far as all decay products are measured except two neutrinos.
Newly introduced cut parameter which define a existence of kinematic solution 
 works to select $B\bar B$ dual semileptonic decay  events.

The neutrino  reconstruction method is 
improved  by using a constraint from  $B$ meson momentum
which is enabled by a classification of  decay particles
into daughters from the tag $B$ meson or from the signal $B$ meson.
in order to reduce time for calculation.
Combination problem is solved  minimizing 
probability density functions with  the simulated annealing  method
in order to reduce time for calculation.
As tag $B$ meson is selected without specifying decay modes,
reconstruction efficiency can be larger and this method has
a statistical advantage against other methods.

\section{ Full reconstruction method}

Once one of $B$ meson is reconstructed in hadronic decay mode,
events are $B\bar B$ from $\Upsilon(4S)$ and 
rest of all particles are assigned to the other $B$.
Though we are not ready yet to show the result
of $b\rightarrow u\ell\nu$ study,
this method provides almost unbiased $B$ decay sample
which is very useful for  inclusive studies.
Here we show the results on inclusive lepton
study from $B$ meson( a study of $|V_{cb}|$ ) as a
superior feature of the full reconstruction.

$B$ meson is reconstructed with following hadronic decay modes,
$$\bar B^0\rightarrow D^{*+}\pi^-, D^{*+}\rho^-,  D^{*+}a_1^-,
D^{+}\pi^-, D^{+}\rho^-, D^{+}a_1^-, $$
$${B^-}\rightarrow D^{*0}\pi^-, D^{*0}\rho^-, D^{*0}a_1^-,
D^{0}\pi^-$$
as a tag side $B$ meson.
We did not include decay modes having larger background events
as further selection in  kinematics is not applicable in
lepton inclusive study.
We obtained 22k $B^0$ tagged events and 25k $B^-$ tagged events
 from $78 fb^{-1}$ on-resonance data and
$8.8 fb^{-1}$ off-resonance data 
taken at Belle~\cite{BELLE} detector with KEKB asymmetric 
collider~\cite{KEKB}.
Number of events are estimated from the fitting of
$\Delta E$( Energy difference) distribution after selecting on
$M_b$( beam-constrained $B$ meson mass ) distribution.
Lepton signal is checked from the rest of tracks with momentum
larger than  $0.6 GeV/c$.
Merits of this method are that  the flavor and  charge  of
the signal $B$ meson is known from the tagged $B$ meson,
which enable to obtain inclusive branching fractions for
$B^0$ and $B^-$ meson  separately and
inclusive lepton spectra from $B$ meson rest-frame,
not from $\Upsilon(4S)$ rest-frame.

Preliminary results of inclusive lepton spectra for $B^0$ and 
$B^-$ meson
and for electron and muon are measured  after background
subtraction
and corrections.
Averaged lepton inclusive branching fraction is
${\cal B}(B\rightarrow X_c \ell\nu) = 0.1119 \pm 0.0020_{exp.} \pm 
0.0031_{theo.}$, 
which is consistent with the previous results~\cite{bxlv}.
The ratio of charged $B$ to neutral $B$ inclusive
branching fraction is
${\cal B}(B^+\rightarrow X_c \ell\nu)/
{\cal B}(B^0\rightarrow X_c \ell\nu) = 1.14 \pm 0.04_{exp.} \pm
0.01_{theo.}$,
which is consistent with the life time ratio of $B$ meson.
This branching fraction is converted into $|V_{cb}|$
using the formula based on the operator product expansion(OPE) and
the heavy quark expansion(HQE)~\cite{PDG} and result in
$|V_{cb}| = ( 4.13 \pm 0.07_{exp.} \pm 0.25_{theo.} )\times 10^{-2}$.

Application of the full reconstruction to the $b\rightarrow u\ell\nu$
is in progress and will be shown in near future.

\section{$D^{(*)}\ell\nu$ tag}

The second method is using $D^{(*)}\ell\nu$ tag.
As these semileptonic modes are the most established decay modes in $B$
meson decay and have  large branching fractions, 
these are good  for $B$ meson tag. 
So we handle two $B$ mesons  decaying like
$B_{tag}\rightarrow D^{(*)}\ell\nu$ and $B_{sig}\rightarrow X_u\ell\nu$.
Separation of two $B$ mesons are automatically done if $B_{tag}$ 
is specified and 
rest of particles except signal lepton in the event are assigned to $X$ system
in the signal side.
But we have to treat two neutrinos in the final state which is 
non-trivial thing.

\begin{figure}
\begin{center}
\epsfxsize=10cm
\epsfbox[0 0 463 307 ]{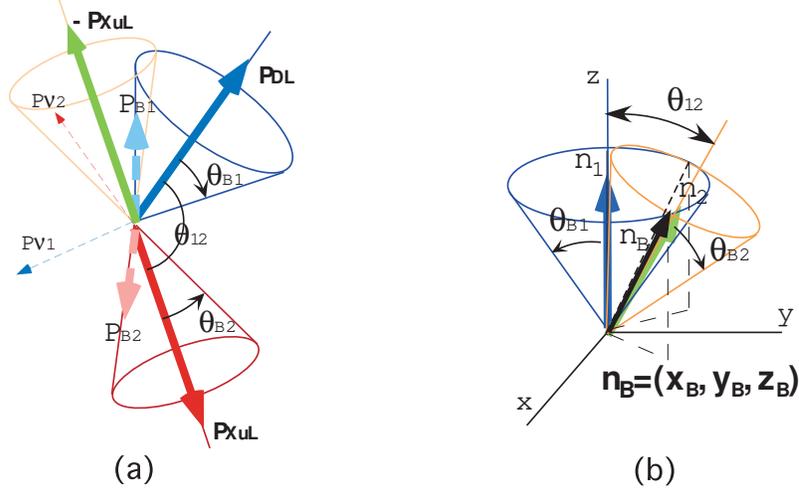}
\end{center}
%\begin{flushleft}
\caption{ Kinematics of two $B$ mesons decaying into semileptonic 
modes( left figure ) and the frame definition in solving the
real $B$ meson direction( right figure ).\hfill
}
%\end{flushleft}
\label{fig:kinema}
\end{figure}

In the single $B$ meson which decays into the semileptonic mode like
$B\rightarrow D\ell\nu$, 
we observe $D$ system and $\ell$ and also  
we know the absolute momentum of $B$ meson 
and its mass at $B$-factory experiment on  $\Upsilon(4S)$.
An assumption of  zero-mass for neutrino derives 
the angle between observed $D\ell$ 
system and $B$ meson's direction($\theta_{B_1}$), 
which means the true $B$ meson direction is on the surface of the cone 
defined with angle $\theta_{B_1}$ around the  direction of $D\ell$ system.
The missing degree of freedom(DOF) in this kinematics is only one and 
corresponds to the azimuthal angle of $B$ direction.
When we extend this situation to two $B$ decays from $\Upsilon(4S)$ 
and both $B$ meson
decay into semileptonc modes 
($B_1\rightarrow D\ell\nu$ and $B_2\rightarrow X\ell\nu$), 
we have two missing DOF in total
but they are filled out from the constraint that
two $B$ mesons move back-to-back  each other at $\Upsilon(4S)$ 
rest frame. 
This implies the real $B$ direction is on the cross lines of two cones
defined by each $B$ semileptonic decay systems when one of $B$ meson decay 
system is spatially inverted.
If the unit vector of $D\ell$ and $X\ell$ are defined as $n_1(0,0,1)$ and 
$n_2(0,\sin\theta_{12},\cos\theta_{12})$ ($\theta_{12}$ shows angle 
between $D\ell$ and $X\ell$ system), 
real $B$'s directional vector is $n_B(x_B,y_B,z_B)$ where 
$z_B=\cos\theta_{B_1}$, 
$y_B=(\cos\theta_{B_2}-\cos\theta_{B_2}\cos\theta_{12})/\sin\theta_{12}$ and
$$x_B=\pm\sqrt{1-{1\over \sin\theta_{12}}(\cos^2\theta_{B_1} +
\cos\theta^2_{B_2} - 2\cos\theta_{B_1}\cos\theta_{B_2}\cos\theta_{12})}.$$
If the assumption of two $B$ decay into semileptonic modes is correct and 
all decay products are observed except two neutrinos, 
$x_B^2$(inside of square root of $x_B$) must be 
greater than 0 and this provides severe selection condition 
of both $B$ semileptonic decay events.

\subsection{Validity of $D\ell\nu$ tag}

The validity of this method is checked using $B\rightarrow X_c\ell\nu$ decays
in signal side where tag side is also decaying to $D^{(*)}\ell\nu$.
This sample is obtained 
from $78 fb^{-1}$ data ( corresponding to 85 million $B\bar B$ events )
just requiring $X$ system contains a single 
charged $K$ in 
addition to existence of $D^{(*)}$ and oppositely charged lepton pairs 
whose momentum exceed $1.0 GeV/c$
where rest of particles are assigned to $X$ system, total charge of event 
is zero
and $x^2_B > -0.2$.
Neutral energy is required to be less than $300 MeV$.
Clear and sharp $D$ and $D^*$ peaks are observed in the $M_X$ distribution
 and smaller tails are very consistent between data and MC which is 
normalized by 
number of $B\bar B$ events, 
where MC doesn't include $b\rightarrow u$ components.
Lepton spectrum is also consistent between data and MC.
This method can select $X_c\ell\nu$ events with high purity and good 
$M_X$ resolution.

\subsection{$B\rightarrow X_u\ell\nu$}

\begin{figure}
\begin{center}
\epsfxsize=14cm
\epsfbox[0 0 680 330 ]{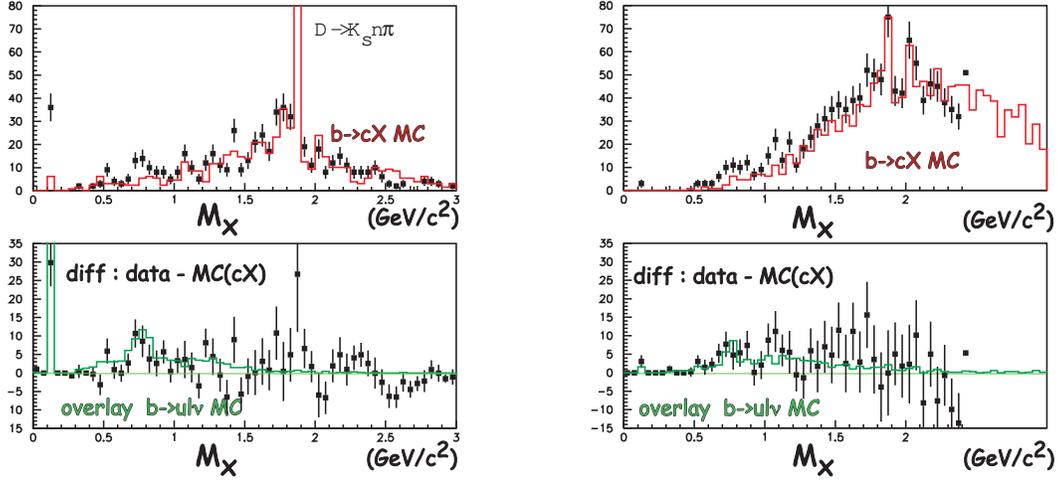}
\end{center}
\caption{ The $M_X$ distributions( solid square for data and histogram
for generic MC ) for the all-charged mode after all
requirements are shown in the top figure. 
The
background from $B\rightarrow X_c\ell\nu$ subtracted $M_X$
distribution is shown in the bottom. 
Solid squares show for data and histogram shows $u\ell\nu$ MC based on
ISGW2. 
Right figures show the same for the $\pi^0$-associated mode.
}
\label{fig:mxu}

\end{figure}

In order to select $X_u\ell\nu$ we require $X$ system doesn't contain 
charged $K$, the other requirements are same as for $|V_{cb}|$ 
sample selection.
When total neutral activity in $X$ system is less than  $300 MeV$,
events are assigned to all-charged mode.
When the neutral energy exceeds $300 MeV$ and  $\pi^0$s are reconstructed 
from neutral clusters, reconstructed $\pi^0$ is attached to $X$ system 
and the event is assigned  to $\pi^0$-associated mode. 
Rest of events are discarded as poorly measured events
to keep good $M_X$ resolution and S/N.
Fig.~\ref{fig:mxu} show $M_X$ distribution of 
$B\rightarrow X_u\ell\nu$ candidates
for all-charged and $\pi^0$ associated modes.
Sharp peaks at $D$ and $D^*$ are attributed from 
$D^{(*)}\rightarrow K^0_S n\pi; K^0_S\rightarrow \pi^+\pi^-$
as we didn't remove $K^0_S$. 
Data show clear excesses in $M_X$ distribution  comparing to 
$b\rightarrow c$ MC distribution at $\pi\ell\nu$ in all-charged mode and 
at $\rho\ell\nu$ in both modes.
If the signal region of $M_X$ is defined  below $1.5 GeV/c^2$,
we obtain $82 \pm 19$ excess in all-charged mode and $92 \pm 21$ 
in $\pi^0$-associated mode, where statistical error includes both data and MC
statistics.
In exclusive decay mode, we observe $30 \pm 6.4$ events in 
$\pi^-\ell^+{\nu}$ and $28 \pm 18 $ events (depend on non-resonant shape)
in $\rho^0\ell^+\nu$.

\subsection{Branching Fraction}

In order to obtain the branching fraction and $V_{ub}$, corrections are
very important and largely depend on models we choose.
Corrections are divided into two parts, one is a detection efficiency 
and the other is a correction of  extrapolation to unobserved 
kinematic regions.
We use 7 different models to investigate two corrections.
The exclusive model(ISGW2)~\cite{isgw2} 
and the hybrid model combining exclusive 
model below $\omega$ in $M_X$  and 
inclusive model~\cite{OPE}~\cite{inclusive}  
above it
are chosen for the detection efficiency. 
Correction factor of extrapolation  is estimated by
inclusive models with several sets of b-quark mass and $\mu^2_\pi$ 
values.
The maximum differences are assigned to systematic error of these corrections.

Branching fraction is calculated by
$${\cal B}={1\over 2}\Sigma {N_{obs}(M^i_X)-N_{BG}(M^i_X)\over \epsilon(M^i_X)}
{1\over \epsilon_{ext}}{1\over 2N_{B\bar B}},$$
where 
$N_{obs}$ and $N_{BG}$ show observed and $b\rightarrow c$ MC expected 
number of events, 
$\epsilon(M^i_X)$ shows a detection efficiency of i-th $M_X$ bin  
and $\epsilon_{ext}$ expresses a correction for the extrapolation.
Obtained branching fraction is \\
${\cal B}(B\rightarrow X_u\ell\nu) = ( 2.62 \pm 0.63_{stat.} 
\pm 0.23_{sys.}
\pm 0.05_{b\rightarrow c} \pm 0.41_{b\rightarrow u} )\times 10^{-3}.$

\begin{figure}
\begin{center}
\epsfxsize=5cm
\epsfbox[0 0 235 255 ]{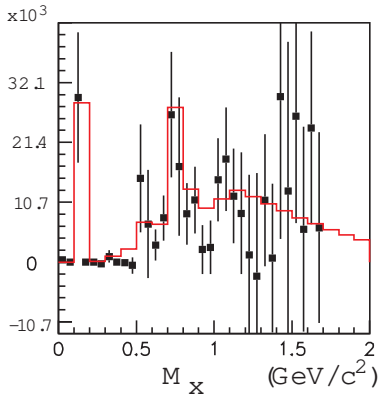}
\end{center}
\caption{ The efficiency corrected $M_X$ distribution
for data( solid square with error ) and for the hybrid model( histogram ).
}
\label{fig:cormxu}
\end{figure}

\section{Advanced neutrino reconstruction}

For the standard neutrino reconstruction method,
a neutrino is reconstructed as a missing momentum
in the final state and
it cannot distinguish $B\bar B$ events from continuum events.
This advanced method separate decay daughters into
the signal side and tag side using the probability density function
defined  by various kinematic variables.
There are a bunch of combinations to separate decay products into two
categories and it is not practical to try all of combinations.
The annealing technique~\cite{anneal} is introduced to obtain a solution of the
combination problem within  a  reasonable iteration.
After classification of decay daughters, $B$ meson constraints can be
used to calculate kinematic variables such as the $B$ meson momentum,
$B$ meson energy and so on, which improve the signal to noise ratio
 significantly.

\begin{figure}
\begin{center}
\epsfxsize=14cm
\epsfbox[0 0 515 225 ]{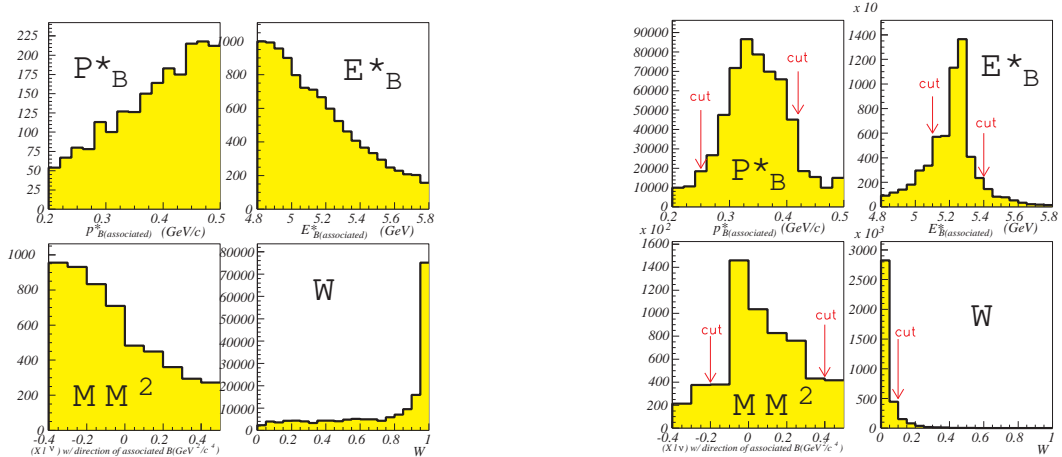}
\end{center}
\caption{ The density probability functions for 
$B$ meson momentum in CM-rest frame( $p^*_B$ ), $B$ meson energy
in the rest frame( $E^*_B$ ), the missing mass square for neutrino( $MM^2$ )
and the incorrect probability for the random sample ( before minimization ) 
are shown in the left  and the same figures after minimization are shown 
in the right. 
}
\label{fig:anneal}
\end{figure}

The annealing method is a algorithm to solve the combination problem.
We define the incorrect probability for a solution using the probability
density functions for a correctly separated sample and randomly
separated one by MC events where probability density functions are
defined by several kinematic distributions.
$$W = {{\bf{\it PDF}}(random; p^*_B, E^*_B, MM^2, ... )\over{
  {{\bf{\it PDF}}(correct; p^*_B, E^*_B, MM^2, ... ) +
{\bf{\it PDF}}(random; p^*_B, E^*_B, MM^2, ... )}}}$$
Search for the optimal combination is carried out by minimizing $W$
 through changing a paricle from signal $B$ category to the tag side
category and vice versa.
With a certain probability a assignment of a particle is not taken even
it provides better $W$ in order to prevent a stack into the local minimum.
Fig.~\ref{fig:anneal} show several kinematic variables before minimization and after.

After interation of this process, the optimal combination for two $B$
meson is obtained. 
We select events whose $W$ is consistent with zero, obtained 
$B$ meson candidate is consistent with
the real $B$ meson using $P^*_B$ and $E^*_B$  
and missing mass calculated with $B$ meson constraint  is 
consistent with neutrino mass.

\subsection{Validity of this method}

This new method is checked using  two different control  samples extracted 
from real data and MC data. 
$M_X$ distribution is studied using $B\rightarrow D^*\ell\nu$ sample 
which has been selected requiring a existance of $D^*$ from 
the mass difference of $M(D\pi)$ and $M(D)$ and $\cos\theta_{B-D^*\ell}$ and
$D*\ell\nu$ is considered as the signal $B$ meson.
Processing this neutrino reconstruction scheme to the sample, two $B$ mesons 
are separated and the most probable $B(\rightarrow X\ell\nu)$ meson 
is selected which must be same as the $B$ meson tagged in sample selection.  
Observed $M_X$ spectrum has a peak at $M_{D^*}$ and the shape is consistent
between data and MC except for its normalization( $\epsilon(data)/\epsilon(MC)
=0.902\pm0.040$ ). 
$q^2$ distribution is also checked using $B\rightarrow J/\Psi X$ sample
where one of leptons from $J/\Psi$ decay is treated as neutrino.
Obtained $q^2$ distribution shows a peak at $q^2=M^2_{J/\Psi}$ and
same shape for data and MC, though the normalization between data and MC 
is slightly different like in the $M_X$ distribution from $D*\ell\nu$ samples
( $\epsilon(data)/\epsilon(MC)=0.862\pm0.046$ ).
Normalization differences between data and MC are considered to be from 
uncertainty of hadronic decay mode in $B$ meson decay table of MC generation 
and this effect($\epsilon(data)/\epsilon(MC)=0.886\pm0.034$ )
 is corrected in the $X_u\ell\nu$ analysis.  
%figure??

\subsection{Branching fraction of $B\rightarrow X_u\ell\nu$}

\begin{figure}
\begin{center}
\epsfxsize=14cm
\epsfbox[0 0 732 451 ]{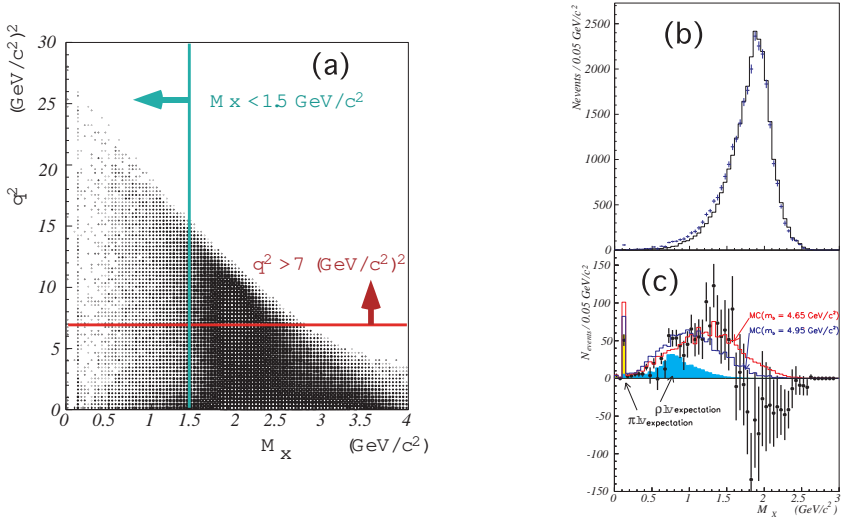}
\end{center}
\caption{ $M_X$ vs. $q^2$ distribution for selected sample( a ), 
the projected $M_X$ distribution after requirement of $q^2 < 7 GeV^2 $ 
for data( cross ) and MC( histogram )
( b ) and the background subtracted $M_X$ distribution ( c ) are shown.
In the last figure, two solid histgrams show expected distributions 
from inclusive models with two different $m_b$ ( $m_b = 4.95 GeV/c^2$ and    
$m_b = 4.65 GeV/c^2$ ) and shaded histograms show expected distributions from 
ISGW2 model for $\pi\ell\nu$ and $\rho\ell\nu$.   
}
\label{fig:mxu2-an}
\end{figure}

Through this advanced neutrino reconstruction, 
$B\rightarrow X\ell\nu$ candidate are selected 
from 85 Million $B\bar B$ sample corresponding to $78 fb^{-1}$.
Selected sample  shown in fig.~\ref{fig:mxu2-an}-(a) 
contains huge $X_c\ell\nu$ components 
which are considered as  backgrounds for 
$B\rightarrow X_u\ell\nu$ study.
The  kinematic regions for the signal are limtted 
in $M_X < 1.5 GeV/c^2$ and 
$q^2 > 7 GeV^2/c^2$ in order to reduce $X_c$ bacgrounds and 
to optimize the signal sensitivity.
We  obtaine 1148 events excess 
to the background in these region.
The branching fraction of $B\rightarrow X_u\ell\nu$ is obtained from
$$B\rightarrow X_u\ell\nu = {1\over 2}{(N_{obs} - a\times N_{bkg})\over
{a\times \epsilon_{ff}}} {1\over{2N_{B\bar B}}},$$
where $N_{obs}$ and $N_{bkg}$ is obtained number of events 
in the signal region for data and MC respectively
$a=\epsilon(data)/\epsilon(MC)$ and $\epsilon_{ff}$ shows a detection 
efficiency of signal estimated from MC.
Models used for efficiency estimations are same as the previous method.
Obtained branching fraction is \\
${\cal B}(B\rightarrow X_u\ell\nu) = ( 1.64 \pm 0.14_{stat.} 
\pm 0.36_{sys.}
\pm 0.28_{b\rightarrow c} \pm 0.22_{b\rightarrow u} )\times 10^{-3}.$

\section{Extraction of $|V_{ub}|$ and systematics}

We used the following formula~\cite{PDG} to extract $|V_{ub}|$ from
${\cal B}(B\rightarrow X_u\ell\nu)$,
$$|V_{ub}|= 0.00445\times({{\cal B}(b\rightarrow u\ell\nu)\over{0.002}}
{1.55 ps\over{\tau_b}})^{1/2}\times( 1 \pm 0.020_{OPE} \pm 0.052_{m_b} ),$$
where $\tau_b=1.606 \pm 0.012$ is used as a average of 
$B^+$ and $B^0$ lifetime~\cite{PDG}. 
Extracted $|V_{ub}|$ are very prepliminary and  
$|V_{ub}| 
= ( 5.00 \pm 0.60_{stat.} \pm 0.23_{sys.} 
\pm 0.05_{b\rightarrow c} \pm 0.39_{b\rightarrow u} \pm 0.36_{theo.})
\times 10^{-3}  $( for $D^*\ell\nu$ ),
$|V_{ub}| = ( 3.96 \pm 0.17_{stat.} \pm 0.44_{sys.} 
\pm 0.34_{b\rightarrow c} \pm 0.26_{b\rightarrow u} \pm 0.29_{theo.})
\times 10^{-3} $( for advanced $\nu$ recon. ).
Obtained values are consistent each other within error as well as results
from other measurements.
Systematic errors from several items are considered and  
listed in table~\ref{table:table1} for each method.

\begin{table}
\caption{Systematics of ${\cal B}(B\rightarrow X_u\ell\nu)$
for $D^{(*)}\ell\nu$ tag and $\nu$ reconstruction method.
  }
\label{table:table1}
\begin{center}
\hspace{9.5cm}
\begin{tabular}{ll|ll}
\hline
$D^{(*)}\ell\nu$ tag & \% error  &  adv. $\nu$ recon. & \% error  \\ \hline
Tracking 	&  6     &  B.G. fraction/signal eff.  & 18 \\
$\pi^0$ recon.  &  3     &  $K_L$ contamination  & 2 \\
recon. eff.     &  3.2     &  lepton ID  	& 12 \\
K ID		& 2	& lepton fake		& 1 \\
lepton ID	& 4	& $M_X$ shape 		& 2 \\
Normalization	& 2.2	& non-$B\bar B$ contami. & $< 1$ \\
 \hfill [ sub total  & 9	&	& 22 ]  \\ \hline
$b\rightarrow c$ related & 	& $B\rightarrow X_c\ell\nu$  &  \\
consistency $M_X < 1.8 GeV/c^2 $ & 2.1 &  model dependence & 17	\\ \hline
$b\rightarrow u\ell\nu$ &    &$b\rightarrow u\ell\nu$ &  \\
recon. and extrapolation & 15.6  & 		& 13 \\
 ( recon.  only		& 5.0 )   &		&  \\ \hline
\end{tabular}
\end{center}
\end{table}

\section{Conclusion}

We present preliminary results on $B\rightarrow X_u\ell\nu$ branching 
fraction using two new methods.\\
~~- for $D^*\ell\nu$ \\
~~~~~${\cal B}(B\rightarrow X_u\ell\nu) = ( 2.62 \pm 0.63_{stat.} 
\pm 0.23_{sys.}
\pm 0.05_{b\rightarrow c} \pm 0.41_{b\rightarrow u} )\times 10^{-3},$\\
~~~~~$|V_{ub}| 
= ( 5.00 \pm 0.60_{stat.} \pm 0.23_{sys.} 
\pm 0.05_{b\rightarrow c} \pm 0.39_{b\rightarrow u} \pm 0.36_{theo.})
\times 10^{-3} $\\
~~- for advanced $\nu$ recon.\\
~~~~~${\cal B}(B\rightarrow X_u\ell\nu) = ( 1.64 \pm 0.14_{stat.} 
\pm 0.36_{sys.}
\pm 0.28_{b\rightarrow c} \pm 0.22_{b\rightarrow u} )\times 10^{-3},$\\
~~~~~$|V_{ub}| = ( 3.96 \pm 0.17_{stat.} \pm 0.44_{sys.} 
\pm 0.34_{b\rightarrow c} \pm 0.26_{b\rightarrow u} \pm 0.29_{theo.})
\times 10^{-3} $\\
Bothe methods provide good $S/N$  in addition to good $M_X$ resolution for 
$D^{(*)}\ell\nu$ tag and good efficiency for the advanced neutrino
reconstruction tag.

\end{document}